# CORRELATIONS BETWEEN QUARKS, MONOPOLES AND INSTANTONS[*]

S. THURNER, H. MARKUM and W. SAKULER

*Institut für Kernphysik, TU Wien, Wiedner Hauptstraße 8-10*
*A-1040 Vienna, Austria*
E-mail: thurner@ds1.kph.tuwien.ac.at

## ABSTRACT

We analyze topological objects in pure QCD in the presence of external quarks by calculating the distributions of instanton and monopole densities around static color sources. We find a suppression of the densities close to external sources and the formation of a flux tube between a static quark–antiquark pair. The similarity in the behavior of instantons and monopoles around static sources might be due to a local correlation between these topological objects. On an $8^3 \times 4$ lattice at $\beta = 5.6$, it turns out that topological quantities are correlated approximately two lattice spacings.

## 1. Introduction

One of the crucial questions in the context of understanding the confinement mechanism in QCD is the condensation of color magnetic monopoles. In the picture of the dual Meissner effect, the QCD vacuum is considered as a coherent state of color magnetic monopoles which force the color electric field between a quark–antiquark pair into an Abrikosov flux tube, leading to a linear confinement potential.[1] However there are problems in deciding whether the condensation actually occurs or not. It is therefore interesting to calculate to what extent monopoles are correlated with other topological quantities, in particular with topological charges.

In the present work we focus on the distributions of monopole densities and topological charge densities around static quarks and mesons which are obtained by computing correlation functions between Polyakov lines and the local values for the corresponding topological quantities. We observe a similar behavior of both correlations and thus we investigate correlation functions between topological objects. First results of the size of the correlations between color magnetic monopoles and topological charges are reported.

## 2. Theory

In order to investigate monopole currents one has to project $SU(3)$ onto its abelian degrees of freedom, such that an abelian $U(1) \times U(1)$ theory remains.[2] This aim can be achieved by various gauge fixing procedures. We employ the so-called maximal

---

[*]Supported in part by "Hochschuljubiläumsstiftung der Stadt Wien" under Contract No. H217/94

abelian gauge which is most favorable for our purposes. After gauge fixing pure QCD may be regarded as a theory of color charges and color magnetic monopoles. A gauge transformation of a gauge field element $U(x,\mu)$ is given by

$$\tilde{U}(x,\mu) = g(x)U(x,\mu)g^\dagger(x+\hat{\mu}) \ , \tag{1}$$

where $g(x) \in SU(3)$. The maximal abelian gauge is imposed by maximizing the functional

$$R = \sum_{x,\mu,i} |\tilde{U}_{ii}(x,\mu)|^2. \tag{2}$$

To extract abelian parallel transporters one has to perform the decomposition

$$\tilde{U}(x,\mu) = c(x,\mu)u(x,\mu) \ , \tag{3}$$

with

$$\begin{aligned} u(x,\mu) &= \text{diag}\left[u_1(x,\mu), u_2(x,\mu), u_3(x,\mu)\right] \ , \\ u_i(x,\mu) &= \exp\left[i \arg \tilde{U}_{ii}(x,\mu) - \frac{1}{3}i\phi(x,\mu)\right] \ , \\ \phi(x,\mu) &= \sum_i \arg \tilde{U}_{ii}(x,\mu)\Big|_{\text{mod } 2\pi} \in (-\pi,\pi]. \end{aligned} \tag{4}$$

Since the maximal abelian subgroup $U(1) \times U(1)$ is compact, there exist topological excitations. These are color magnetic monopoles which have integer-valued magnetic currents on the links of the dual lattice:

$$m_i(x,\mu) = \frac{1}{2\pi} \sum_{\Box \ni \partial f(x+\hat{\mu},\mu)} \arg u_i(\Box) \ , \tag{5}$$

where $u_i(\Box)$ denotes a product of abelian links $u_i(x,\mu)$ around a plaquette $\Box$ and $f(x+\hat{\mu},\mu)$ is an elementary cube perpendicular to the $\mu$ direction with origin $x+\hat{\mu}$. The magnetic currents form closed loops on the dual lattice as a consequence of monopole current conservation. Finally the local monopole density is given by

$$\rho(x) = \frac{1}{3 \cdot 4V_4} \sum_{\mu,i} |m_i(x,\mu)| \ . \tag{6}$$

For the implementation of the topological charge on the lattice there exists no unique discretization. In this work we restrict ourselves to the so-called field theoretic definitions which approximate the topological charge in the continuum [3]

$$q(x) = \frac{g^2}{32\pi^2} \epsilon^{\mu\nu\rho\sigma} \text{Tr}\left(F_{\mu\nu}(x)F_{\rho\sigma}(x)\right) \ , \tag{7}$$

in the following ways:

$$q_L^{(P,H)}(x) = -\frac{1}{2^4 32\pi^2} \sum_{\mu\nu\rho\sigma=\pm 1}^{\pm 4} \tilde{\epsilon}_{\mu\nu\rho\sigma} \text{Tr } O_{\mu\nu\rho\sigma}^{(P,H)}(x) \ , \tag{8}$$

with

$$O^{(P)}_{\mu\nu\rho\sigma}(x) = U(x,\mu)U(x+\hat{\mu},\nu)U^\dagger(x+\hat{\nu},\mu)U^\dagger(x,\nu)$$
$$\times\ U(x,\rho)U(x+\hat{\rho},\sigma)U^\dagger(x+\hat{\sigma},\rho)U^\dagger(x,\sigma)\ , \qquad (9)$$

for the plaquette prescription and

$$O^{(H)}_{\mu\nu\rho\sigma}(x) = U(x,\mu)U(x+\hat{\mu},\nu)U(x+\hat{\mu}+\hat{\nu},\rho)U(x+\hat{\mu}+\hat{\nu}+\hat{\rho},\sigma)$$
$$\times\ U^\dagger(x+\hat{\nu}+\hat{\rho}+\hat{\sigma},\mu)U^\dagger(x+\hat{\rho}+\hat{\sigma},\nu)U^\dagger(x+\hat{\sigma},\rho)U^\dagger(x,\sigma)\ , \quad (10)$$

for the hypercube prescription. The lattice and continuum versions of the theory represent different renormalized quantum field theories, which differ from one another by finite, non-negligible renormalization factors.[4] A simple procedure that enables one to get rid of renormalization constants, while preserving physical information contained in lattice configurations, is the cooling method. The cooling procedure systematically reduces quantum fluctuations, and suppresses differences between the different definitions of the topological charge. In our investigation we have employed the so-called "Cabbibo–Marinari method".

A static quark is represented by the Polyakov loop $L(\vec{r})$ which describes the propagation of a charge with infinite mass

$$L(\vec{r}) = \frac{1}{3}\text{Tr}\prod_{t=1}^{N_t} U(\vec{r},t;\mu=4)\ , \qquad (11)$$

where $N_t$ is the temporal extension of the lattice. To measure the distribution of topological quantities around a static quark–antiquark pair we calculate correlation functions

$$\langle L(0)L^\dagger(d)\rho(r)\rangle\ , \qquad \langle L(0)L^\dagger(d)q^2(r)\rangle \qquad (12)$$

for several quark–antiquark separations $d$; they are normalized to the corresponding cluster values $\langle L(0)L^\dagger(d)\rangle\langle\rho\rangle$ and $\langle L(0)L^\dagger(d)\rangle\langle q^2\rangle$. Since topological objects with opposite sign are equally distributed, we correlate the absolute value of the monopole and the square of the instanton densities.

## 3. Results

Our simulations were performed on an $8^3\times 4$ lattice with periodic boundary conditions using the Metropolis algorithm. The observables were studied in pure QCD at inverse gluon coupling $\beta = 6/g^2 = 5.6$ employing the $SU(3)$ Wilson plaquette action for the gluons. We made 50000 iterations and measured our observables after every 50th iteration. Each of these 1000 configurations was first cooled and then subjected to 300 gauge fixing steps enforcing the maximal abelian gauge. The potentials and the quantities concerning the topological charge according to the plaquette prescription were determined before gauge fixing, the abelian observables afterwards.

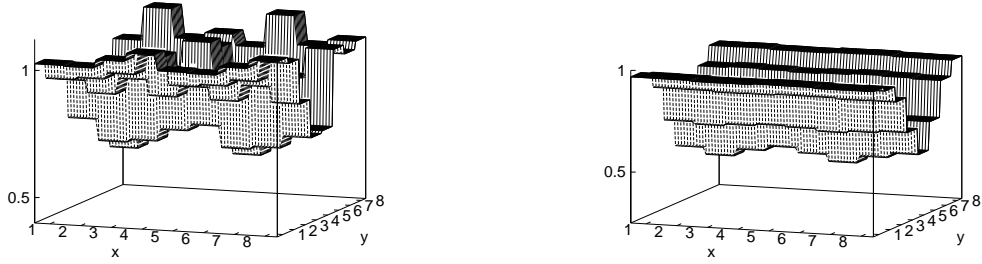

Fig. 1. Color magnetic monopole density (l.h.s.) and topological charge density after 5 cooling steps (r.h.s.) around a static meson with $q\bar{q}$-separation $d = 4$.

The distributions of color magnetic monopoles and topological charges around a static meson with $q\bar{q}$-separation $d = 4$ according to Eq. (12) are shown in Fig. 1. A local suppression of the topological quantities around the charges is clearly visible. The valley of reduced density reflects the formation of a flux tube between the static $q\bar{q}$-pair.

In Fig. 2 the correlation functions between topological quantities $\langle \rho(0)\rho(d)\rangle$, $\langle q^2(0)q^2(d)\rangle$ and $\langle \rho(0)q^2(d)\rangle$ are given; they are normalized after the subtraction of their cluster values. One observes an influence of cooling on the correlations. The original $q^2q^2$-correlation is $\delta$-peaked and its range grows rapidly with cooling. In contrast the $\rho\rho$-correlation decreases with cooling. The $\rho q^2$-correlation seems rather insensitive to cooling and clearly extends over more than two lattice spacings, indicating some nontrivial local correlation between monopoles and topological charges.

## 4. Discussion and Outlook

We have studied the distributions of color magnetic monopoles and topological charges in the presence of static color sources.[5] Motivated by a qualitatively very similar behavior of those quantities we have looked at correlation functions amongst them. We found that our monopole and instanton correlations have a range of about two lattice spacings. We presented the first indications for a nontrivial relation between monopoles and instantons. There is an enhanced probability that at the locations of monopoles also instantons can be found on gauge average. One might speculate whether a natural many–body gluon–state like a fundamental condensate exists. An appropriate quasi–particle transformation to a new basis where monopole and instanton states can be expanded could demonstrate their overlap more directly. To test this conjecture, at present we analyze the correlations between color magnetic monopoles and topological charges per gauge field configuration. For a more reliable statement these correlations have to be examined for fixed spacial separation without volume average. Results for the quark–gluon plasma–phase will be reported elsewhere. Sea quarks will be switched on applying the pseudofermionic algorithm. Besides we plan to use a geometric definition of the topological charge for future investigations.

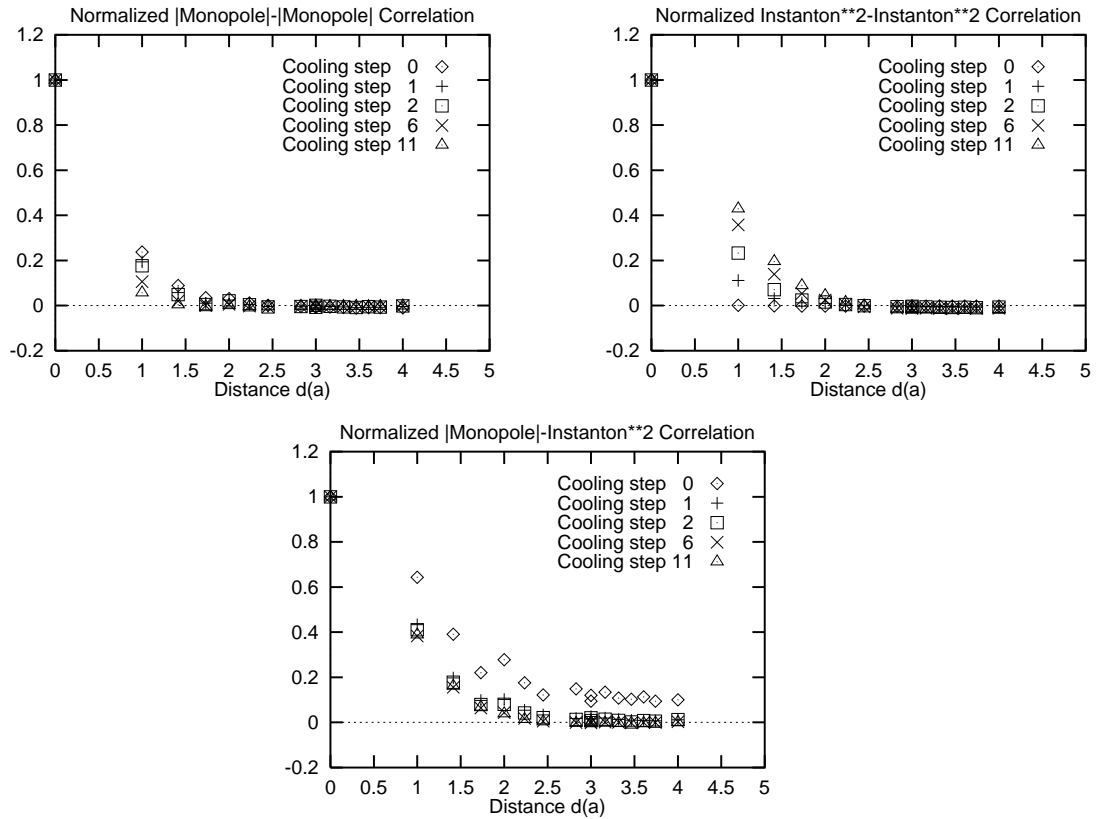

Fig. 2. Correlations between the different topological observables for several cooling steps.

1. G. 't Hooft, in *High Energy Physics*, Proceedings of the EPS International Conference, Palermo 1975, ed. A. Zichichi (Editrice Compositori, Bologna, 1976); S. Mandelstam, *Phys. Rep.* **23C** (1976) 245.
2. G. 't Hooft, *Nucl. Phys.* **B190** (1981) 455.
3. P. Di Vecchia, K. Fabricius, G. C. Rossi and G. Veneziano, *Nucl. Phys.* **B192** (1981) 392; *Phys. Lett.* **B108** (1982) 323.
4. M. Campostrini, A. Di Giacomo and H. Panagopoulos, *Phys. Lett.* **B212** (1988) 206; B. Allés, A. Di Giacomo and M. Giannetti, *Phys. Lett.* **B249** (1990) 490.
5. H. Markum, W. Sakuler and S. Thurner, *Nucl. Phys.* **B** *(Proc. Suppl.)* **39B,C** (1995) 235; M. Faber, H. Markum, Š. Olejník and W. Sakuler, *Phys. Lett.* **B334** (1994) 145.